\documentstyle[pra,aps]{revtex}
\begin{document}
\title{S-wave Scattering Length 
       from Effective Positronium-Positronium Interaction
       for Bose-Einstein Condensates}
\author{K. Oda, T. Miyakawa, H. Yabu and T. Suzuki}
\address{Department of Physics, Tokyo Metropolitan University, 
         Hachioji, Tokyo 192-0397, Japan}
\date{\today}
\maketitle
%
%
\begin{abstract}
The s-wave scattering length 
for the Positronium-Positronium interaction
is estimated semi-phenomenologically 
with the long-range van~der Waals force and 
the short-range repulsive potential that represents the hard core 
between two positronium. 
The obtained value of the scattering length is 
$a \sim 0.44{\rm\,nm}$, 
and its stability is also checked for different parametrizations. 
Using this value, the Gross-Pitaevskii equation can be fixed 
for the Positronium Bose-Einstein condensates (Ps BEC). 
The static properties of Ps BEC 
are studied from the solutions of that equation. 
The phase-transition temperature shift of the Ps BEC 
due to the Ps-Ps interaction is also evaluated 
with the mean-field approximation. 
\end{abstract}
%
%
\section{Introduction}
Recent realizations of Bose-Einstein condensates 
for trapped Alkali-atom gas 
have made a great development in quantum physics of many-body system 
where the condensation phenomenon is one of the most important concepts 
and used in a vast area of physics 
such as atomic, laser, condensed-matter, nuclear, 
elementary-particle physics 
and also in cosmology\cite{bec}.  
After the experimental success of the condensation, 
a lot of physical phenomena has been observed and studied on Alkali-atom BEC 
that include collective excitations, sound propagations, 
quantum interference, etc\cite{review}.

The success of the Alkali-atom BEC and 
the developments of the laser cooling and trapping techniques 
encouraged the study for the BEC of bose/fermi particles 
other than Alkali atoms:  
recently, the BEC of Rb isotope 
with the sympathetic cooling technique\cite{symp}
and the potassium fermi condensates have been observed\cite{fermi}. 
The bose/fermi mixed condensates are also studied 
for the potassium isotopes\cite{mixed,miyakawa}. 

The positronium (Ps), the bound state of the electron and the positron,
has been also an interesting candidate for the BEC: 
\begin{itemize}
\item
The Ps is the lightest atom with the mass $m_{\hbox{Ps}} \sim 2 m_e$ 
where $m_e$ is the electron mass: $10^3$ times lighter than the hydrogen atom. 
It leads to the large value of the critical temperature $T_C$ 
of the Bose-Einstein phase transition: 
in free-gas and continuous approximation, 
$T_C$ becomes 10 times higher than that of hydrogen BEC 
for the trapping potential of same size ($T_C \propto m^{-1/3}$). 
The effect of the Ps-Ps interaction for $T_C$ will be discussed 
in this paper. 
\item
The Ps has finite life-time and self-annihilates into photons: 
$Ps \to 2\gamma$ with $\tau_{\hbox{pPs}} \sim 0.125 {\rm\,ns}$ 
for para-Ps and
$Ps \to 3\gamma$ with $\tau_{\hbox{oPs}} \sim 142 {\rm\,ns}$ 
for ortho-Ps. 
Thus, the Ps BEC should be an unstable condensate and decays 
with gamma-ray radiations, 
the coincident observation of which will give 
the precise measurements for position/momentum distributions of the BEC. 
Those radiated gamma-ray photons are also suggested to show 
new physical phenomena such as gamma-ray laser 
by spontaneous amplifications and the supperradiance. 
\end{itemize}
However, 
it should be noted that these very short life-times will make 
technical difficulties in experimental Ps BEC formations:  
$\tau_{\hbox{Ps}}$ are in the comparable order with the relaxation time 
$\tau_{rel} \sim {\rm ns}$ in atomic phenomena 
so that the cooling methods that depend on the natural relaxation 
(like the evaporative cooling) might be less effective 
for the Ps BEC formation. 
At present, several faster cooling methods are planned 
by several experimental groups for Ps-BEC formations, 
and the preliminary experiments are progressing\cite{liang,platzman,hirose}. 

Originally, the Ps BEC was considered as $e^+$-$e^-$ condensates 
as a source of gamma-ray laser\cite{baldwin,liang,bertolotti}, 
the $e^+$-$e^-$ plasma in the strong magnetic field 
at the neutron star surface\cite{varma} and the storage ring\cite{winterberg}, 
and also as the source of the gamma-ray burst in astro-physics\cite{ramaty}. 
The Ps BEC itself has been suggested 
as a source of the gamma-ray laser\cite{loeb}, 
but detailed studies for the properties of Ps BEC 
including the Ps-Ps interaction effects have not been done. 

In the binding structure, 
the Ps is similar with the hydrogen atom (H), 
where the main differences are mass 
of positively charged particle ($m_p/m_{e^+} \sim 2000$) 
(and their stability). 
The BEC of poralized H has been reported with $N \sim 10^9$ atoms 
\cite{fried}.   
More similar object are excitons in semiconductors: 
the bound states of electron and hole in valence band 
with $m_h/m_e \sim 0.7$. 
The exciton has also finite life-time $\tau \sim 30{\rm\,ns}$ 
and decays into phonon, 
so that it is quite similar with the Ps.  
The BEC of excitons in CuO has also been reported \cite{lin}. 
The experimental success of those BEC encourages strongly 
studies of Ps BEC.  

In this paper, 
we derive the Ps-Ps interaction potential
with the long-range van~der Waals interaction 
and the short-range cut-off that represents the hard core repulsion 
of positronium. 
The s-wave scattering length is estimated from the solution 
of the Schr{\"o}dinger equation using this potential. 
With the given scattering length, 
the Gross-Pitaevski equation for the ground-state wave function 
(order parameter) of Ps BEC is given, 
and, solving this equation with the trapping potential, 
we study the ground-state properties of Ps BEC. 
We also estimate the Ps-Ps interaction effects 
for the critical temperature $T_C$ 
in the mean-field approximation, 
which are shown to be negligible 
except large particle number BEC.   
\section{Effective Ps-Ps Interaction Potential}
\vspace{0.5cm}
\noindent
i) {\it van~der Waals Potential for Ps-Ps Interaction}  
\smallskip

We consider the system of two positronium Ps(1) and Ps(2). 
The spacial coordinates of the electron and positron 
composing Ps(1) are denoted by ${\bf x}_1$ and ${\bf x}_a$ 
and those of Ps(2) are by ${\bf x}_2$ and ${\bf x}_b$, 
with which the relative coordinates are defined by 
${\bf r}_{ij} ={\bf x}_j -{\bf x}_i$. 
Especially, for relative coordinates between electron and positron, 
we denote
\begin{equation}
  {\bf r}_1 \equiv {\bf r}_{a1},  \quad
  {\bf r}_2 \equiv {\bf r}_{b2}. 
\label{eQa}
\end{equation}
The hamiltonian of the system is 
\begin{equation}
  H =H_{CM} +H_1 +H_2 +H_{int}, 
\label{eQb}
\end{equation}
where $H_{CM}$ is the center-of-mass part that can be neglected 
in the derivation of the Ps-Ps interaction 
because of the translational invariance. 
The $H_{1,2}$ in (\ref{eQb}) are the internal parts of Ps(1,2) 
with the Coulomb potential: 
\begin{equation}
     H_{1,2} =\frac{{\bf p}_{1,2}^2}{m_e} -\frac{\alpha \hbar c}{r_{1,2}}, 
\label{eQc}
\end{equation}
where $r_{1,2} =|{\bf r}_{1,2}|$ 
and ${\bf p}_{1,2}$ are the momenta conjugate with ${\bf r}_{1,2}$. 
The $\alpha$ in (\ref{eQc}) is a fine structure constant and 
$\alpha =e^2/(4\pi \epsilon_0 \hbar c) \sim 1/137$ in the SI unit system. 
The interaction part of Ps(1,2) is represented by
\begin{equation}
     H_{int} =\alpha \hbar c 
              \left( \frac{1}{r} +\frac{1}{r_{12}} 
               -\frac{1}{r_{a2}} -\frac{1}{r_{b1}} \right), 
\label{eQd}
\end{equation}
where $r_{ij} =|{\bf r}_{ij}|$ 
and $r \equiv r_{ab} =|{\bf r}_{ab}|$ is the distance between 
two positrons, Ps(1) and Ps(2). 

Let us consider the case 
where the distance between Ps(1,2) are very large: 
$r \equiv r_{ab} \geq r_1, r_2$.  
In that case, the Ps-Ps interaction potential can be determined 
by the van~der Waals potential\cite{london,pauling}. 

Eq. (\ref{eQd}) can be expanded by $1/r$, and we get 
the dipole-dipole interaction: 
\begin{equation}
  H_{int} \approx -\frac{\alpha \hbar c}{r^3}
          \left\{3 \frac{({\bf r}_1 \cdot {\bf r}) 
                    ({\bf r}_2 \cdot {\bf r})}{r^2}  
                    -{\bf r}_1 \cdot {\bf r}_2
    \right\}. 
\label{eQe}
\end{equation}
When $r \gg 0$, the interaction term $H_{int} \sim 1/r^3$ in (\ref{eQe}) 
can be evaluated with perturbation. 
The unperturbed part is just $H_0 =H_1 +H_2$ defined by (\ref{eQc}) 
and the zeroth-order unperturbed wave functions that diagonalize 
the $H_0$ are
$\psi_{nlm}(1) \otimes \psi_{n'l'm'}(2)$ 
where $\psi_{nlm}(i)$ is the eigenstate 
of Ps(i),  
$H_i \psi_{nlm}(i) =E^{(0)}_n  \psi_{nlm}(i)$,  
with the binding energy: 
\begin{equation}
  E^{(0)}_n =- m_e c^2 \alpha^2/ \{4(n+1)^2\}. 
\label{eQf}
\end{equation}
The $n$, $l$, $m$ are principal, azimuthal and magnetic quantum 
numbers. 
In the Ps state function, 
we dropped the spin angular momentum part and spin quantum numbers 
$S=0,1$ and $S_z$, 
because the spin-spin/spin-orbital interactions are weak, 
and give a very small modifications that can be neglected 
in the present semi-qualitative estimations. 

From the second-order perturbation for $H_{int}$, 
we get the van~der Waals potential: 
\begin{equation}
  E^{(2)} =-\alpha^2 (\hbar c)^2 \frac{6}{r^6} 
            \sum_{n \neq 0,l,m,n' \neq 0,l',m'}
            \frac{|{\langle 0 | d_1 | \psi_{nlm}(1)    \rangle}|^2
	          |{\langle 0 | d_2 | \psi_{n'l'm'}(2) \rangle}|^2}
	         {E_n^{(0)}+E_{n'}^{(0)}
                 -E_0^{(0)}-E_{0}^{(0)}},   
\label{eQg}
\end{equation}
where $d_{1,2} =z_{1,2}$ are dipole-moment operators of Ps(1,2) 
and ${| 0 \rangle} ={| \psi_{000} \rangle}$.  
Using eq. (\ref{eQf}), 
the denominator of eq. (\ref{eQg}) becomes 
\begin{equation}
  {E_n^{(0)}+E_{n'}^{(0)} -E_0^{(0)}-E_{0}^{(0)}} 
   =\frac{\alpha \hbar c}{4a_0} [2 -(n+1)^{-2} -(n'+1)^{-2}], 
\label{eQh}
\end{equation}
where 
$a_0 =\hbar c/(\alpha m_e c^2) \sim 0.0529{\rm\,nm}$ is the Bohr radius. 
The term $(n+1)^{-2}$ in the denominator of eq. (\ref{eQh}) is very small 
for large value of $n$,   
so that it can be neglected in the first approximation 
\begin{equation}
  {E_n^{(0)}+E_{n'}^{(0)} -E_0^{(0)}-E_{0}^{(0)}} 
   \sim \frac{\alpha \hbar c}{2 a_0}, 
\label{eQi}
\end{equation}
Using eq. (\ref{eQi}) and the completeness condition: 
$\sum_{n,l,m} {\langle 0 | d_1 | \psi_{nlm} \rangle} 
              {\langle \psi_{nlm} | d_1 | 0 \rangle}
 ={\langle 0 | z^2 | 0 \rangle}$,
eq. (\ref{eQg}) becomes 
\begin{equation}
  E^{(2)}=-\frac{6 \alpha \hbar c (2 a_0)}{r^6}
           |\langle 0 | z^2 | 0 \rangle|^2. 
\label{eQj}
\end{equation}
Using ${\langle 0 | z^2 | 0 \rangle} =(2 a_0)^2$ 
with the Coulomb wave function $\psi_{000}$ 
for the ground state, 
we get van~der Waals potential: 
\begin{equation}
  E^{(2)} =-6 \alpha \frac{\hbar c (2 a_0)^5}{r^6}.  
\label{eQk}
\end{equation}

The error for the above approximation (\ref{eQk}) 
has been evaluated by numerical summation for (\ref{eQg}) 
for the H-H interaction \cite{pauling}. 
When their results are rescaled for Ps potential, 
they obtained 6.47 instead of 6 in (\ref{eQk}).  
It suggests that the present results (\ref{eQk}) 
should be a reasonable estimation. 

\vspace{0.5cm}
\noindent
ii) {\it Effective Ps-Ps interaction potential with the short-range hard core}
\smallskip

In the short range part, 
the real part of the Ps-Ps interaction should be characterized 
by repulsive hard core at the distance of atomic diameter 
where two positronium begin to overlap. 
In the present paper, 
we treat this hard core potential phenomenologically 
in two manners: 
A) the cut-off potential at $r =r_C$, $V_{\hbox{Ps-Ps}}^{CO}$ (Fig.~2) and 
B) the Lenard-Jones potential with $r^{-12}$-hard core (Fig.~3): 
\begin{equation}
  V_{\hbox{Ps-Ps}}^{LJ}(r) =\alpha \hbar c \left[-6 \frac{(2 a_0)^5}{r^6} 
                                  +C \frac{(2 a_0)^{11}}{r^{12}}
                            \right],
\label{eQl}
\end{equation}

The phenomenological parameters $r_C$ and $C$ 
in $V_{\hbox{Ps-Ps}}^{CO}$ and $V_{\hbox{Ps-Ps}}^{LJ}$ 
are fixed to reproduce the binding energy $B_{\hbox{Ps}_2}$ 
of the molecular state Ps${}_2$, 
which has not been confirmed experimentally 
but predicted theoretically/numerically 
with the hamiltonian (\ref{eQb},\ref{eQc},\ref{eQd}) \cite{abdel}. 
Recent results for $B_{\hbox{Ps}_2}$ by elaborate numerical calculations 
are around $0.435{\rm\,eV}$ \cite{kinghorn,kozlowski,varga}. 

To reproduce $B_{\hbox{Ps}_2} =0.435{\rm\,eV}$ 
with solving the Schr{\"o}dinger equation for $V_{\hbox{Ps-Ps}}$ numerically, 
we obtain 
A) $r_C =1.78 a_0$ 
(with the potential depth $v_0 =-6.13 \alpha \hbar c/a_0$) 
for $V_{\hbox{Ps-Ps}}^{CO}$, 
and 
B) $C =2.25$ 
($v_0 =-2.0 \alpha \hbar c/a_0$ and 
$r'_C =1.90 a_0$: the distance where $V_{\hbox{Ps-Ps}}(r'_C) =0$)
for $V_{\hbox{Ps-Ps}}^{LJ}$. 
It should be noticed that two phenomenological potentials 
have almost the same hard-core distances 
$r_C, r'_C \sim 1.8 a_0$,  
which are consistent with the diameter of Ps, $2 a_0$. 

We should comment about the effect of Ps spin state. 
Because of the Pauli principle, 
the Ps${}_2$ bound state exists in singlet state, 
so that the effective Ps-Ps potential 
using the binding energy of Ps${}_2$ as an input parameter 
should be considered for singlet channel. 
However, because the $r_C$ of $V_{\hbox{Ps-Ps}}^{CO,LJ}$ is very close to 
$2 a_0$ that should be expected to be the hard core length 
for nonsinglet channel \cite{platzman}, 
the present potential should be applicable also for these channels. 

\section{s-wave Scattering Length by Effective Ps-Ps Interaction}
For the Ps-Ps s-wave scattering, 
the relative wave function of two positronium becomes 
$\Psi({\bf r}) =r \chi(r)$ that satisfies the Sch{\"o}dinger equation: 
\begin{equation}
  -\frac{\hbar^2}{2 \mu} 
   \frac{d^2}{dr^2} \chi(r) 
  +V_{\hbox{Ps-Ps}}(r) \chi(r) =E \chi(r), 
\label{eQm}
\end{equation}
where $\mu$ is a reduced mass $\mu =m_{\hbox{Ps}}/2 \sim m_e$. 

The s-wave scattering length $a$ is obtained from the asymptotic form 
of the $E=0$ solution for eq. (\ref{eQm}) as 
\begin{equation}
  \chi(r \to \infty) \sim 1 -\frac{r}{a}.  
\label{eQn}
\end{equation}

In fig.~4, 
we show the numerical results for the $E=0$ wave function $\chi(r)$ 
of eq. (\ref{eQm}) with $V_{\hbox{Ps-Ps}}^{CO,LJ}$ defined 
in the previous section. 
The one peak at $r \sim 3.5 a_0$ shows 
that (only) one weak bound state exists for this potential 
and it corresponds to the Ps${}_2$ molecular state 
whose binding energy has been used as an input for $V_{\hbox{Ps-Ps}}$. 
As seen in fig.~4, for $r \gtrsim 5 a_0$, 
$\chi(r)$ in fig.~4 are in the asymptotic linear region, 
so that the scattering length $a$ can be read off 
by $\chi(a) =0$: 
\begin{equation}
  \hbox{A) } a =0.440{\rm\,nm} \quad\hbox{for $V_{\hbox{Ps-Ps}}^{CO}$},  \quad
  \hbox{B) } a =0.437{\rm\,nm} \quad\hbox{for $V_{\hbox{Ps-Ps}}^{LJ}$}.   
\label{eQo}
\end{equation}
Both results are almost consistent, 
and it means that the s-wave scattering length 
(the low-energy limit)  
do not depend on the detailed form of the short-range potential. 
Existence of only one weakly bound state in Ps-Ps system 
is also essential for the rigidness of $a$ given 
with the effective potential

For the illustration of the effects by bound states 
for the scattering length, 
the $s$-wave scattering state function $\chi(r)$ 
is shown in fig.~5
for the effective interaction potential fitted with 
H-H binding energy.  
That potential has  the larger $v_0$ than the Ps-Ps potential 
and produces many excited molecular states 
that can be seen as the oscillating behavior of $\chi(r)$ 
for $r \lesssim 5 a_0$. 
Its behavior depends on the short-range part of the potential 
given only in phenomenological manner 
in the effective potential, 
so that it leads to the impreciseness 
of the connected long-range behavior of $\chi(r)$ 
from which the scattering length $a$ is given.
For the Ps-Ps case, 
it should also be too rough approximation 
to treat the molecular bound state Ps${}_2$ itself
as that of the effective potential given in this paper, 
but the long-range part of the scattering state 
should be more reliable 
because its short-range behavior is very simple 
and the potential has only one bound state 
that was used for the input parameter.
The consistency between the scattering lengths 
given with $V_{\hbox{Ps-Ps}}^{CO}$ and $V_{\hbox{Ps-Ps}}^{LJ}$ 
also supports it. 

Currently-used $s$-wave scattering lengths for Alkali atoms 
are, 
$5.77{\rm\,nm}$ for ${}^{87}$Rb \cite{boesten}, 
$2.75{\rm\,nm}$ for ${}^{23}$Na \cite{tiesinga} and 
$-1.45{\rm\,nm}$ for ${}^{7}$Li \cite{abraham}. 
The resultant Ps-Ps $s$-wave scattering lengths in (\ref{eQo}) 
are $a \sim 0.44{\rm\,nm}$, 
which is smaller than those for Alkali atoms in one order 
but larger than that for polarized H atom (triplet), 
$a =0.065{\rm\,nm}$ \cite{jamieson}. 
It shows that the Ps-gas behaves more like perfect gas 
than the Alkali atom gas. 
The rough estimation for the Ps-Ps scattering length can also be 
made directly from the binding energy of Ps${}_2$ \cite{wu}: 
$a \sim (2 m_e B_{\hbox{Ps}_2})^{-1/2} \sim 0.3{\rm\,nm}$ \cite{platzman}, 
which is also consistent with the present result.  

\section{Ps BEC with Ps-Ps Interaction}
The dynamical behavior of BEC is represented 
by the ground-state wave function (order parameter) 
$\psi(x)$ 
and, for weak-interacting bose gas 
in the harmonic oscillator trapping potential at $T=0$, 
it satisfies the Gross-Pitaevski (GP) equation \cite{review}: 
\begin{equation}
  i \hbar \frac{\partial \psi}{\partial t} 
    = -\frac{\hbar^2}{2m_{\hbox{Ps}}} \nabla^2\psi
      +\frac{1}{2} m_{\hbox{Ps}} \omega_{HO}^2 r^2 \psi 
      +g N |\psi|^2 \psi, 
  \label{eQp}
\end{equation}
where $\omega_{HO}$ is the angular frequency 
of the harmonic oscillator potential, 
and $N$ is the condensed bose particle number: 
the normalization condition for $\psi$ is taken as
\begin{equation}
  \int{d^3x} |\psi(x)|^2 =1. 
\label{eQq}
\end{equation}
For low-density bose gas, 
the boson-boson interaction is dominated by 
the low-energy s-wave scattering, 
and the interaction constant $g$ in (\ref{eQp}) 
is given with the s-wave scattering length $a$ \cite{review}: 
\begin{equation}
  g =\frac{4 \pi \hbar^2}{m_{\hbox{Ps}}} a.  
\label{eQr}
\end{equation}
Using the dimensionless variables 
${\tilde {\bf r}} ={\bf r}/a_{HO}$, 
${\tilde t} =\omega_{HO} t$ and 
${\tilde \psi} =a_{HO}^{3/2} \psi$ 
with the harmonic oscillator length 
$a_{HO} =\sqrt{\hbar/(m_{\hbox{Ps}} \omega_{HO})}$,  
the GP equation (\ref{eQp}) becomes 
\begin{equation}
  i \frac{\partial {\tilde \psi}}{\partial{\tilde t}} 
    = -\frac{1}{2} {\tilde \nabla}^2 {\tilde \psi}
      +\frac{{\tilde r^2}}{2} {\tilde \psi}
      +{\tilde g} N |{\tilde \psi}|^2 {\tilde \psi}, 
  \label{eQs}
\end{equation}
where the dimensionless constant ${\tilde g} N$ is 
\begin{equation}
  {\tilde g} N =\frac{4\pi a}{a_{HO}}N. 
\label{eQt}
\end{equation}
Thus, the solutions of GP equation are similar 
for the same value of ${\tilde g} N$. 
It should be noticed 
that the interaction effects come into eq. (\ref{eQs}) 
through ${\tilde g} N$-term, 
so that they are more effective 
for large value of ${\tilde g} N$.   

In fig.~6, 
the solutions of eq. (\ref{eQp}) are shown 
with 
$a =0.437{\rm\,nm}$ in (\ref{eQo}) 
and $\omega =3.1{\rm\,sec^{-1}}$,  
for several boson numbers:
$N =2.6 \times 10^7$ (curve 2), 
$2.0 \times 10^8$ (curve 3), 
$7.9 \times 10^8$ (curve 4), 
$2.5 \times 10^9$ (curve 5). 
With $a(Ps) =0.44{\rm\,nm}$ for Ps, 
we obtain
$a(Ps)/a_{HO} =2.16{\rm\,eV^{-1/2}} \sqrt{\hbar \omega_{HO}}$, 
which is much smaller than 
$a(Rb)/a_{HO} =7962{\rm\,eV^{-1/2}} \sqrt{\hbar \omega_{HO}}$ 
for Rb atom:  
this large difference comes 
not only from weak Ps-Ps interaction $a(Ps)/a(Rb) \sim 1/130$ 
but from the mass difference $\sqrt{m_{\hbox{Ps}}/m_{Rb}} \sim 1/281$. 
Because of the small $a/a_{HO}$ for the Ps-Ps interaction, 
for large particle number $N =2.6 \times 10^7$ (curve 2 in fig.~6), 
the calculated $|\psi(r)|$ changes only $\sim 16\%$ from that  
for the ideal gas $a=0$ (curve 1 in fig.~6). 
To realize $70 \%$ change for $|\psi(0)|$, 
the Ps of $N =2.5 \times 10^9$ should condensate (curve 5 in fig.~6). 
For Rb BEC,  
because of large scattering length $a =5.77{\rm\,nm}$, 
$N =6.7 \times 10^5$ is enough 
for realization of the profile 5 in fig.~6. 

For the ideal gas with no interactions, 
the critical temperature $T_C^0$ 
can be calculated in thermodynamical continuum limit \cite{review}: 
\begin{equation}
  k_B T_C^0 =0.94 \hbar \omega_{HO} N^{1/3}, 
\label{eQu}
\end{equation}
where $k_B$ is a Boltzmann constant. 
The particle interaction has an effect to shift 
the critical temperature:  
$T_C =T_C^0 +\delta T_C$. 
The $\delta T_C$ can be evaluated 
with the mean-field and semiclassical approximation \cite{giorgini}: 
\begin{equation}
  \frac{\delta T_C}{T_C^0} =-1.33 \frac{a}{a_{HO}} N^{1/6}. 
\label{eQv}
\end{equation}
For Alkali atom BEC, eq. (\ref{eQv}) is consistent 
with experimentally observed temperature shift \cite{review} 
($\sim 2\%$ shift for Rb BEC of $N \sim 10^4$). 
In the case of Ps BEC, eq. (\ref{eQv}) becomes 
\begin{equation}
  \frac{\delta T_C}{T_C^0} =-2.87 {\rm\,eV}^{-1/2}
  \sqrt{\hbar \omega_{HO}} N^{1/6}. 
\label{eQw}
\end{equation}
It gives very small shift for $T_C$ for Ps BEC, 
and can be neglected in the first approximation 
except BEC of large $N$. 

It should be noted that the concept of $T_C$ is exact 
for equilibrium (thermodynamical) system. 
For real experimental situations, 
the Ps BEC might be performed as nonequilibrium system 
due to the small Ps life-time 
that is almost the same order with relaxation time. 
In that case, the $T_C$ evaluated here should be considered 
as estimation for the energy scale. 

\section{Summary and Discussions}
In the present paper, 
we studied the effective Ps-Ps interaction potentials 
with interpolating the long-range van~der Waals interaction 
and the short-range repulsive hard core.  
The latter was taken phenomenologically 
in two kinds of parametrizations: 
A) sudden cut-off and B) Lenard-Jones (and cut-off) types. 
The Ps-Ps s-wave scattering length was calculated 
from the solutions of the Scr{\"o}dinger equation 
with those effective potentials, 
and we obtained $a \sim 0.44{\rm\,nm}$. 
It is ($2 \sim 10$)-times smaller than that of the Alkali atoms, 
and larger than that of the polarized hydrogen atom. 
The stability of $a$ for the different parametrizations 
of the short-range interaction was also checked. 
Using that value of the scattering length, 
we studied the ground-state wave functions for Ps BEC
with solving the Gross-Pitaevski equation. 
The Ps BEC is shown to be more perfect-gas-like 
than Alkali-atom BEC because of weaker interaction. 
The critical temperature shift $\delta T_C$ 
due to the Ps-Ps interaction was also estimated 
and shown to be very small 
because of the weak interaction and the small mass of Ps atom. 

As commented in the last section, 
the inclusion of the dissipative effects 
to the Gross-Pitaevski equation is essentially important 
for further studies of Ps BEC physics discussed in the introduction. 
It will be discussed in a future publication\cite{oda}.  
    
%

%
%
\noindent
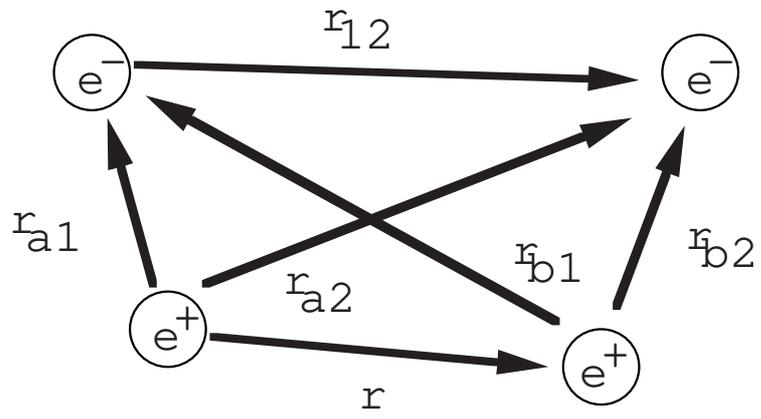
\begin{figure}
\caption{Relative coordinates of $e^+$ and $e^-$ in Ps-Ps system.}
\end{figure}

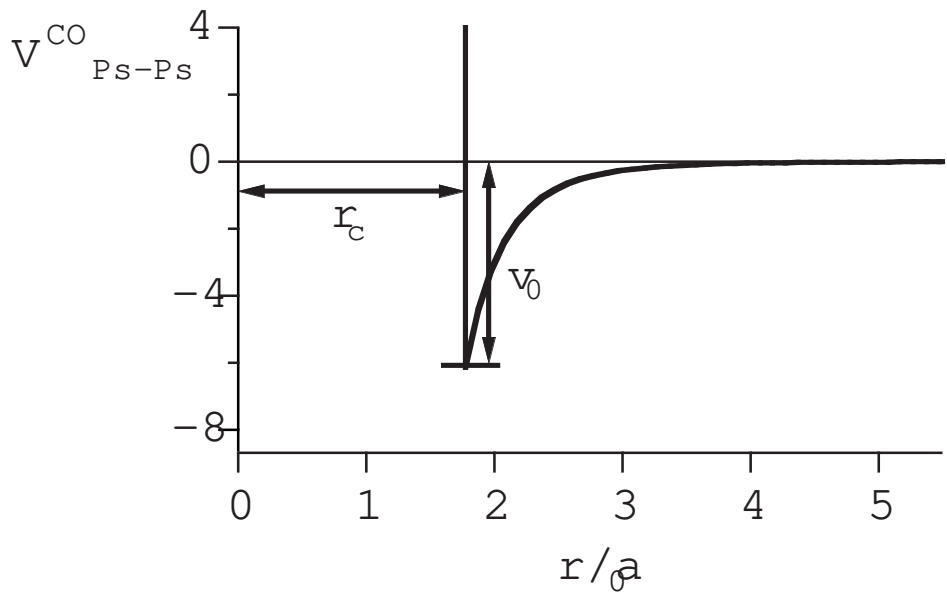
\begin{figure}
\caption{Effective Ps-Ps interaction potential $V_{\hbox{Ps-Ps}}^{CO}$ 
against $r/a_0$. The $a_0 =0.053{\rm\,nm}$ is the Bohr radius.  
The $r_C$ is the cut-off length that represents the hard-core 
repulsive interaction. 
The $v_0$ in figure represents the potential depth.}
\end{figure}

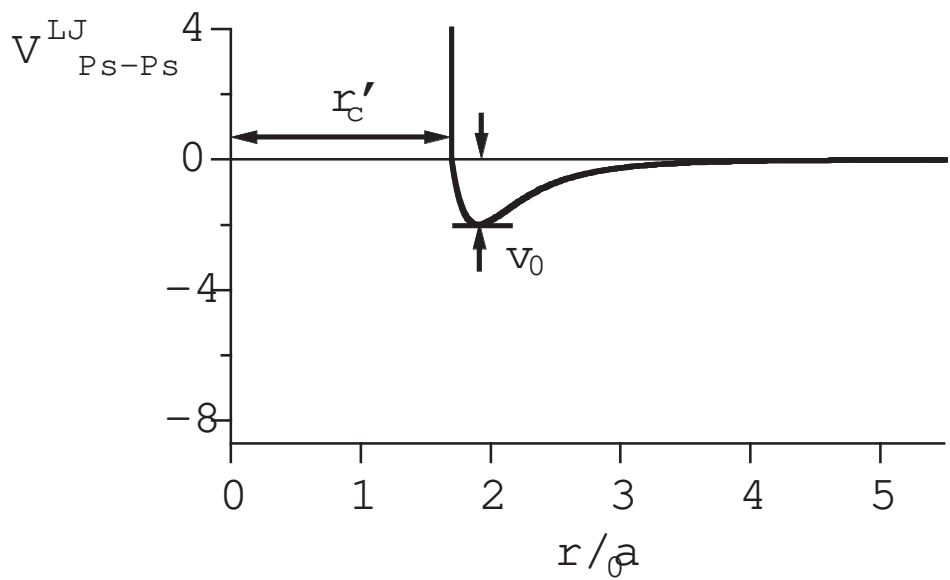
\begin{figure}
\caption{Effective Ps-Ps interaction potential $V_{\hbox{Ps-Ps}}^{LJ}$ 
in unit of $\alpha \hbar c/a_0~ {\rm\,eV}$ 
against $r/a_0$. The $a_0 =0.053{\rm\,nm}$ is the Bohr radius.  
The short-range part is parametrized by the Lenard-Jones-type 
$r^{-12}$-potential. 
The cut-off $r'_C$ is defined by $V_{\hbox{Ps-Ps}}^{LJ}(r'_C) =0$, 
and the $v_0$ in figure represents the potential depth.}
\end{figure}

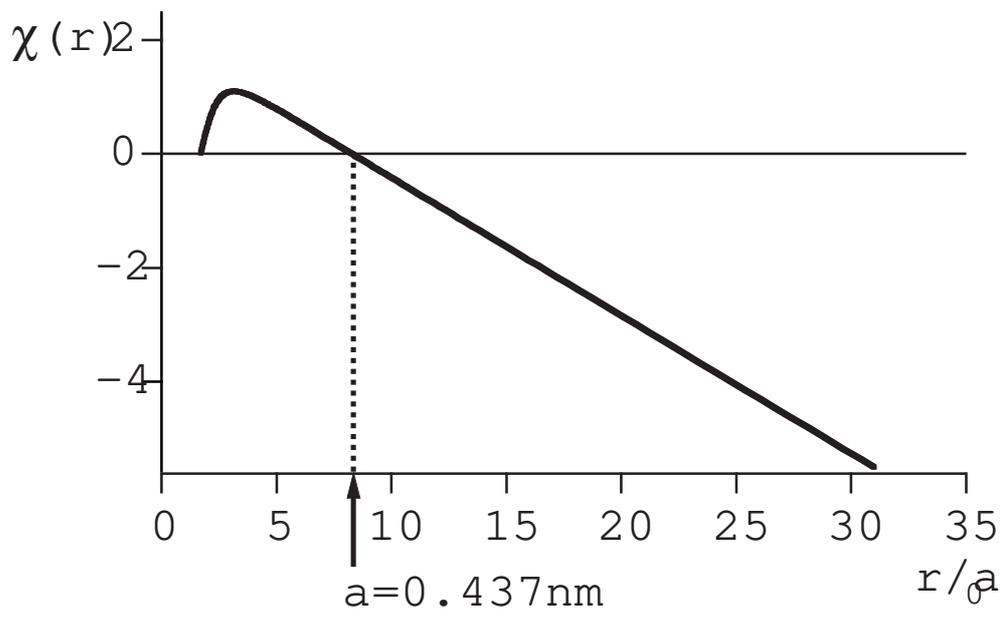
\begin{figure}
\caption{The profile of the s-wave Ps-Ps scattering solution $\chi(r)$ 
for $V_{\hbox{Ps-Ps}}^{LJ}$ in unit of $\alpha \hbar c/a_0~ {\rm\,eV}$ 
against $r/a_0$ ($a_0 =0.053{\rm\,nm}$). 
The s-wave scattering length $a$ can be read off at $\chi(a) =0$, 
and $a =0.437{\rm\,nm}$. 
For $V_{\hbox{Ps-Ps}}^{CO}$, we obtain almost the same result, 
and $a =0.440{\rm\,nm}$.}
\end{figure}

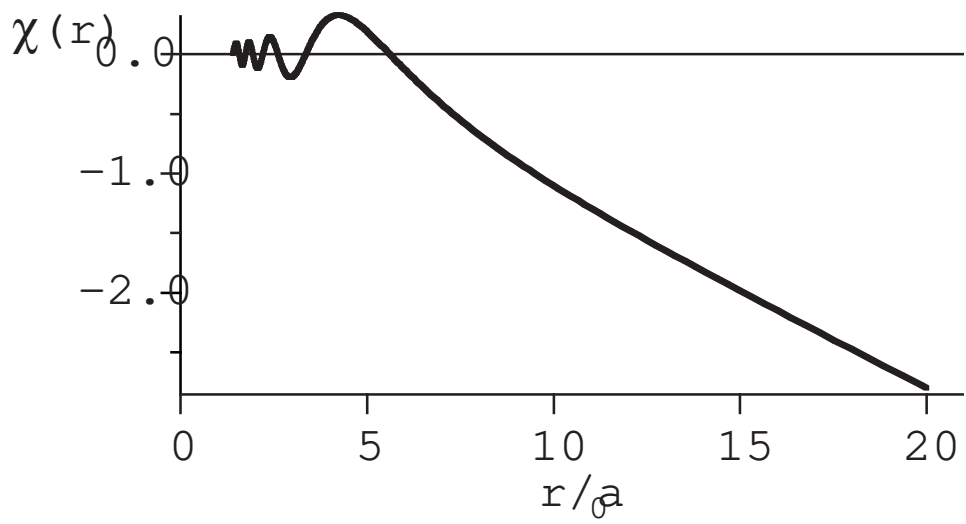
\begin{figure}
\caption{The profile of the s-wave H-H scattering solution $\chi(r)$ 
for $V_{\hbox{Ps-Ps}}^{LJ}$ against $r/a_0$ ($a_0 =0.053{\rm\,nm}$). }
\end{figure}

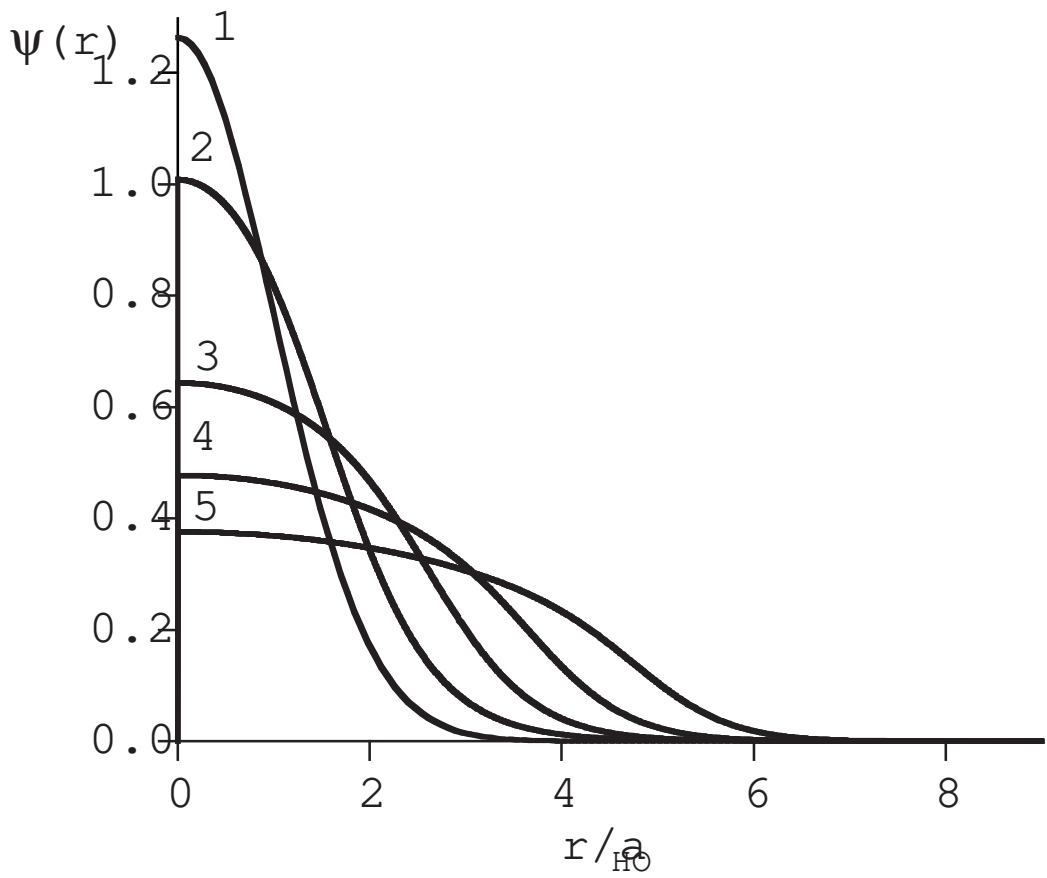
\begin{figure}
\caption{The space-distribution $|\psi(r)|$ for Ps BEC at $T=0$
with the trapping harmonic oscillator potential 
($\omega_{HO} =3.1{\rm\,s^{-1}}$) 
for several condensed Ps numbers: 
$N=2.6 \times 10^7$ (curve 2), 
$2.0 \times 10^8$ (curve 3), 
$7.9 \times 10^8$ (curve 4), 
$2.5 \times 10^9$ (curve 5). 
The curve 1 with gaussian profile corresponds to the BEC 
with no interaction. 
The normalization condition of $\psi$ is given by (\ref{eQq}). }
\end{figure}
%
\end{document}